\documentclass[preprint,showpacs,preprintnumbers,amsmath,amssymb]{revtex4}
\input psfig.sty

\usepackage{graphicx}
\usepackage{dcolumn}
\usepackage{bm}

\begin{document}

\title{Conservative Force Fields in Non-Gaussian Statistics}

\author{J. M. Silva} \email{jmsilva@astro.iag.usp.br}
\affiliation{Instituto de Astronomia, Geof\'{\i}sica e Ci\^encias
Atmosf\'ericas, USP, 05508-900 S\~ao Paulo, SP, Brasil}

\author{R. Silva} \email{raimundosilva@uern.br}
\affiliation{Universidade do Estado do Rio Grande do Norte,
Departamento de F\'{\i}sica - PPGF, 59610-210, Mossor\'o, RN, Brasil
and Universidade Federal do Rio Grande do Norte, Departamento de
F\'{\i}sica - PPGF, 59072-970 Natal - RN, Brasil}

\author{J. A. S. Lima}\email{limajas@astro.iag.usp.br}
\affiliation{Instituto de Astronomia, Geof\'{\i}sica e Ci\^encias
Atmosf\'ericas, USP, 05508-900 S\~ao Paulo, SP, Brasil}

\date{\today}







\begin{abstract}
In this letter, we determine the $\kappa$-distribution function for
a gas in the presence of an external field of force described by a
potential U(${\bf r}$). In the case of a dilute gas, we show that
the $\kappa$-power law distribution including the potential energy
factor term can rigorously be deduced in the framework of kinetic
theory with basis on the Vlasov equation. Such a result is
significant as a preliminary to the discussion on the role of long
range interactions in the Kaniadakis thermostatistics and the
underlying kinetic theory.
\end{abstract}
\pacs{51.10.+y; 05.20.-y; 05.90.+m }

\maketitle


\section {Introduction}

Over the last two decades, a great deal of attention has been paid
to the so-called power-law statistics, both from theoretical and
observational viewpoints. The main motivation is the lack of a
comprehensive and rigorous treatment including gravitational and
Coulombian fields, or more generally, any long range interaction for
which the assumed additivity of the entropy present in the standard
approach is not valid (see, e.g.,
\cite{tsallis1,tsallis2,tsallis3,lima,qkinet1,qkinet2,application}
and references therein).

In this context, the Tsallis
\cite{tsallis1,tsallis2,tsallis3,lima,qkinet1,qkinet2,application}
and Kaniadakis
\cite{k1,k2,kaniadakis05,KaniaH01,rai0506,kaniad2001,scarfone05,kaniadakis04,kania05}
power-law statistics are the most investigated frameworks. In the
former case, the standard Boltzmann-Gibbs formalism is extended
through a new analytic form for the entropy $S_q = k_{\rm{B}} (1 -
\sum_{i}{p_{i}^{q}})/(q - 1)$, where $k_{\rm{B}}$ is the standard
Boltzmann constant, $p_i$ is the probability of the $i$th
microstate, and $q$ is a parameter quantifying the degree of
nonextensivity. This expression has been introduced in order to
extend the applicability of statistical mechanics to system with
long range interactions and has the standard Gibbs-Jaynes-Shannon
entropy as a particular limiting case ($q = 1$). Even the so-called
$q$-nonextensive kinetic theory \cite{lima} has also been developed
and applied for many different contexts ranging from plasmas
\cite{qkinet1} to gravitational systems \cite{qkinet2}.

On the other hand, recent studies on the kinetic foundations of the
$\kappa$-statistics also leads to a power-law distribution function
and a $\kappa$-entropy which emerges naturally in the framework  of
the {\it kinetic interaction principle} (see, for instance, Ref.
\cite{k1,k2,kaniadakis05}). Several physical features  of the
so-called $\kappa$-distribution have also been theoretically
investigated, among them: i) the self-consistent relativistic
statistical theory \cite{kaniadakis05}, ii) the framework of
nonlinear kinetics \cite{KaniaH01} and iii) the H-theorem and
generalization of the chaos molecular hypothesis \cite{rai0506}.
Actually, this $\kappa$-framework leads to a class of one parameter
deformed structures with interesting mathematical properties
\cite{kaniad2001}. In particular, the canonical quantization of a
classical system \cite{scarfone05}, and the so-called Lesche
stability have also been discussed in the $\kappa$-framework
\cite{kaniadakis04}. Still more important, a consistent form for the
entropy (linked with a two-parameter deformations of logarithm
function), which generalizes the Tsallis, Abe and Kaniadakis
logarithm behaviours \cite{kania05} have also been found.

In the experimental front, there also exist some evidence closely
related to the behavior predicted by  the $\kappa$-statistics,
namely, cosmic rays flux, rain events in meteorology
\cite{kaniad2001}, quark-gluon plasma \cite{miller03}, kinetic
models describing a gas of interacting atoms and photons
\cite{rossani04}, fracture propagation phenomena \cite{cravero04},
and income distribution \cite{drag03}, and even the possibility of
improved financial models has also been investigated
\cite{bolduc05}.

Mathematically, the $\kappa$-framework is based on
$\kappa$-exponential and $\kappa$-logarithm functions which are
defined by
\begin{equation}\label{expk}
\exp_{\kappa}(f)= (\sqrt{1+{\kappa}^2f^2} + {\kappa}f)^{1/{\kappa}},
\end{equation}
\begin{equation}\label{expk1}
\ln_{\kappa}(f)= ({f^{\kappa}-f^{-\kappa})/2\kappa},
\end{equation}
\begin{equation}
\ln_{\kappa}(\exp_{\kappa}(f))=\exp_{\kappa}(\ln_{\kappa}(f))\equiv
f.
\end{equation}
The $\kappa$-entropy associated with this $\kappa$-framework is
given by
\begin{equation}\label{e1}
S_{\kappa}(f)=-\int d^{3}p f [\frac{f^{\kappa} -
f^{-\kappa}}{2\kappa}],
\end{equation}
which recovers standard Boltzmann-Gibbs entropy,
$S_{\kappa=0}(f)=-\int f \ln f d^3 p$, in the limit
$\kappa\rightarrow 0$ (see Ref. \cite{k1,k2} for details).

The so-called Kaniadakis entropy reads \cite{k1,abe2004}
\begin{equation}\label{first}
S_\kappa = - \int d^3 p f\ln_\kappa f =  - \langle{\ln_\kappa
(f)\rangle}.
\end{equation}
while the equilibrium velocity distribution can be written as
\cite{k1,k2,kaniadakis05,rai0506}
\begin{equation}
\label{e5} f_0(v) = {A_\kappa}
\left[\sqrt{1+\kappa^2\left(-{mv^2\over 2k_BT}\right)^2} + \kappa
{\left(-{mv^2\over 2k_B T}\right)}\right]^{1 \over \kappa}.
\end{equation}
In this expression $k_B$ is the Boltzmann constant while the
$\kappa$ index denotes a continuous parameter associated to the gas
entropy, and whose main effect at the level of the distribution
function is to replace the standard Gaussian form by a power law.
The quantity $A_\kappa$ is a normalization constant fixed by the
total number of particles in a given volume. As it should be
expected, the above expression reduces to the standard Maxwellian
distribution in the limit $\kappa=0$ for which $A_0= (m/2\pi k_B
T)^{3/2}$.

In this letter we explore how the potential energy term can
rigorously be introduced in the $\kappa$-distribution \cite{1}. More
precisely, we deduce an analytical expression for the equilibrium
distribution of a dilute gas under the action of a conservative
force field with basis on the  stationary solution of the
collisionless Boltzmann equation. As we shall see, this result is
significant as a preliminary to the discussion of long range
interactions according to Kaniadakis thermostatistics and the
underlying kinetic theory.

\section{Vlasov equation and the Boltzmann Factor}

In this section we discuss briefly the standard case, i.e., the
kinetic description of a classical gas under stationary conditions
and immersed in a conservative force field, ${\bf F}=-\nabla U(r)$.
Typical examples are a gas in the earth's gravitational field and
ions in an external magnetic field \cite{Huang,KT}. This kind of
problem is important on their own because it permits to understand
how the molecular motion is modified by force-fields different from
those exerted by the containing vessel or even by the other
particles of the gas. As widely known, its distribution function
differs from the Maxwellian velocity statistics trough an extra
exponential factor involving the potential energy whose general form
reads
\begin{equation}\label{e1}
f({\bf r},v)=n_0\left({m\over 2\pi k_B T} \right)^{3/2}\exp
\left(-{{1\over 2}m v^2 +U({\bf r})\over k_B T}\right),
\end{equation}
where $m$ is the mass of the particles, $T$ is the temperature and
$n_0$ is the particle number density in the absence of the external
force field. In addition, since this distribution function is
normalized, it is easy to see that the number density is given by
\begin{equation}\label{e2}
n({\bf r})=n_0 \exp \left[-{{U({\bf r}) \over k_B T}}\right],
\end{equation}
where the factor, $\exp[-{U({\bf r})/k_B T}]$, which is responsible
for the inhomogeneity of $f({\bf r},v)$, is usually called the
Boltzmann factor. Expression ($\ref{e1}$) follows naturally from an
integration of the collisionless Boltzmann's equation
\begin{equation}\label{e3}
\frac{\partial{\it f}}{\partial{\it t}} + {\bf
v}\cdot\frac{\partial{\it f}}{\partial{\bf r}} + \frac{\bf
F}{m}\cdot\frac{\partial{\it f}}{\partial{\bf v}} = 0,
\end{equation}
when stationary conditions are adopted along with the assumption
that the total distribution can be factored
\begin{equation}\label{e4}
f({\bf r},v)=f_0(v)\chi({\bf r}),
\end{equation}
where $f_0(v)$ represents the Maxwell equilibrium distribution
function, and $\chi({\bf r})$ is a scalar function of ${\bf r}$. As
one may show, after a simple normalization, the resulting expression
for $\chi({\bf r})$ is exactly the Boltzmann factor for the
potential energy of the external field, namely:

\begin{equation}\label{e2a}
\chi({\bf r})=\exp \left[-{{U({\bf r}) \over k_B T}}\right],
\end{equation}
and combining this result with equation (\ref{e4}) the Boltzmann
stationary distribution (\ref{e1}) is readily obtained.

\section{Vlasov Equation and Kaniadakis Kinetic Theory}

Let us now consider a spatially inhomogeneous dilute gas supposed 
in nonequilibrium stationary state at temperature $T$. The gas
is immersed in a conservative external field in such a way that
$f(r,v)d^3v d^3r$ is the number of particles with velocity lying
within a volume element $d^3 v$ about ${\bf v}$ and positions lying
within a volume element $d^3 r$ around ${\bf r}$. In principle, this
distribution function must be determined from the $\kappa$-type
Boltzmann equation:
\begin{equation}\label{e7}
{\bf v}.\nabla_r f - \frac{1}{m}\nabla_r U.\nabla_v f = C_k(f)
\end{equation}
where $C_{\kappa}$ denotes the k-collisional term. The
left-hand-side of the above equation is just the total time
derivative of the distribution function. Hence, the effects
appearing from the $\kappa$-approach can be explicitly incorporated
only through the collisional term. In particular, this means that
the Vlasov equation, or the stationary Boltzmann equation takes the
standard form (see Eq. (7))

\begin{equation}\label{e7a}
{\bf v}.\nabla_r f - \frac{1}{m}\nabla_r U.\nabla_v f = 0.
\end{equation}

In order to introduce the $\kappa$-statistics effects we first
recall that the factorizability condition as given by (\ref{e4}) is
modified in this extended framework. This means that the standard
starting assumption must be extended. In the spirit of the
$\kappa$-formalism, a consistent $\kappa$-generalization of
(\ref{e4}) is
\begin{equation}
\label{e8} f({\bf r} ,v) = A_\kappa \exp_{\kappa}\left[
\ln_{\kappa}\left(\frac{f_0}{A_k}\right) + \ln_{\kappa} \chi({\bf
r})\right],
\end{equation}
where $f_0$ denotes the $\kappa$-velocity distribution and the
constant normalization  has been introduced for mathematical
convenience, and the functions $\exp_\kappa(f)$, $\ln_\kappa(f)$,
were previously  defined by Eqs. (\ref{expk}) and (\ref{expk1}).

Note that in the limit ${\kappa \rightarrow 0}$ the above identity
reproduces the usual properties of the exponential and logarithm
functions. In addition, since $\exp_\kappa(\ln_\kappa f) =
\ln_\kappa (\exp_\kappa(f)) = f$, the standard factored
decomposition (\ref{e4}) is readily recovered in the limit
$\kappa=0$. In what follows, the properties of $\kappa$-exponential
and $\kappa$-log differentiation
\begin{eqnarray}\label{e11}
{d\ln_\kappa f \over dx}= \left({f^{\kappa-1}+f^{-(\kappa+1)}\over 2} \right){df \over dx}, \\
{d \exp_\kappa(f) \over dx}={\exp_\kappa(f)\over \sqrt{1+\kappa^2
f^2}}{df \over dx},
\end{eqnarray}
will also be extensively used. In particular, for the total
$\kappa$-distribution (\ref{e8}), we obtain

\begin{eqnarray*}
\label{e12a} \nabla_{\bf r} f(r,v) =
\frac{\exp_{\kappa}[\ln_{\kappa}f_{0}(v)+\ln_{\kappa}\chi(r)]}
{\exp_{\kappa}^{\kappa}[\ln_{\kappa}f_{0}(v)+\ln_{\kappa}\chi(r)]}\nabla_{r}\ln_{\kappa}\chi(r)\times
\end{eqnarray*}
\begin{equation}
\left\{1+\frac{\kappa\left(\ln_{\kappa}\chi(r)-\frac{mv^{2}}{2k_{B}T}\right)}
{[1+\kappa^{2}(\ln_{\kappa}f_{0}(v)+\ln_{\kappa}\chi(r))^{2}]^{1/2}}\right\}
\end{equation}

\begin{eqnarray*}
\label{e12b} \nabla_{\bf v} f(r,v) =
\frac{\exp_{\kappa}[\ln_{\kappa}f_{0}(v)+\ln_{\kappa}\chi(r)]}
{\exp_{\kappa}^{\kappa}[\ln_{\kappa}f_{0}(v)+\ln_{\kappa}\chi(r)]}\left(-\frac{mv}{k_{B}T}\right)\times
\end{eqnarray*}
\begin{equation}
\left\{1+\frac{\kappa\left(\ln_{\kappa}\chi(r)-\frac{mv^{2}}{2k_{B}T}\right)}
{[1+\kappa^{2}(\ln_{\kappa}f_{0}(v)+\ln_{\kappa}\chi(r))^{2}]^{1/2}}\right\}
\end{equation}

Now, substituting $\nabla_r f$ and $\nabla_v f$ given above into the
stationary Boltzmann equation (\ref{e7}), and performing the
elementary calculations one obtains

\begin{equation}
\label{e13} \nabla_{\bf r} \ln \chi \cdot d{\bf r}= -
\frac{1}{k_BT}\nabla U(r) \cdot d{\bf r}
\end{equation}
the solution of which is
\begin{equation}\label{e14}
\chi({\bf r})= \exp_\kappa \left( - \frac{U({\bf r})}{k_BT} + C
\right),
\end{equation}
where $C$ is an arbitrary constant.

\begin{figure}[t]
\vspace{.2in}\centerline{\psfig{figure=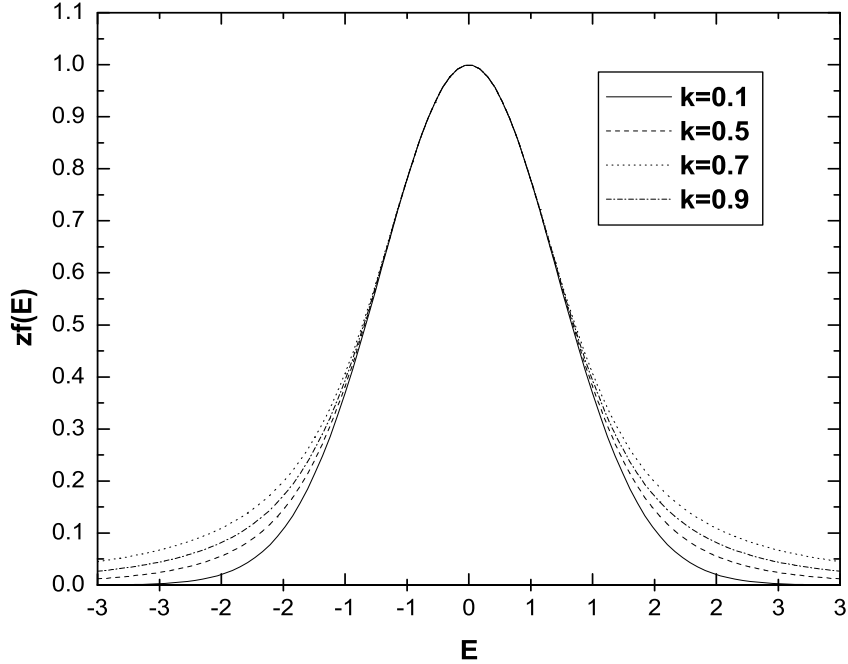,width=5.5truein,height=4.5truein}
\hskip 0.1in} \caption{The $\kappa$-velocity distribution function
$f_{\kappa}(\bf{r},v)$ for typical values of the $\kappa$ parameter
on the interval [-1,1]($z={A_{\kappa}}^{-1}$). In contrast to
Tsallis statistics \cite{tsallis1,tsallis2}, this power law
distributions does not {\it exhibit a thermal cutoff}, that is, a
maximum value allowed for the velocity of the particles. The
$\kappa$-distribution also presents a striking mathematical
property, namely, $f_{\kappa}(\eta)=f_{-\kappa}(\eta)$.}\label{gra2}
\end{figure}

Here, inserting (\ref{e14}) into (\ref{e8}), and integrating the
result in the velocity space it follows that
\begin{equation}
\label{e15} \int A_\kappa
\exp_\kappa\left[\ln_\kappa\left(z{f_0}\right) - \frac{U}{k_BT} + C
\right] d^3v = n({\bf r}).
\end{equation}
Finally, by substituting the expression of $f_0$ and considering a
region where $U({\bf r}) =0$, one finds
\begin{equation}\label{e16}
A_k\int \exp_\kappa\left(- \frac{m{\bf v}^2}{2k_BT} + C \right) d^3v
= n_0,
\end{equation}
and from normalization condition, $n_0 = \int f_0(v) d^3v$, it
follows that the unique allowed value for the integration constant
is $C=0$. Consequently, (\ref{e14}) becomes
\begin{equation}\label{e17}
\chi ({\bf r})= \exp_\kappa\left[- \frac{U({\bf r})}{k_BT}\right],
\end{equation}
which is the $\kappa$-generalized Boltzmann factor.

Finally, by inserting this result into (\ref{e8}), we obtain the
complete $\kappa$-distribution function in the presence of an
external field
\begin{eqnarray*}\label{eq18}
f_{\kappa}({\bf r},v) = A_\kappa\left[\sqrt{1
+\kappa^2\left(-\frac{m{\bf v}^2}{2k_BT} - \frac{U({\bf
r})}{k_BT}\right)^2} + \kappa \left(-\frac{m{\bf v}^2}{2k_BT} -
\frac{U({\bf r})}{k_BT}\right) \right]^{1/\kappa}
\end{eqnarray*}
\begin{equation}\label{later}
\equiv A_\kappa\exp_\kappa(-E/k_B T),
\end{equation}
where $E$ is the total energy of the particles. It thus follows that
a generalized $\kappa$-exp factor for Kaniadakis' thermostatistics
can exactly be deduced if the standard approach is slightly
modified. In Fig. 1 we show $zf(E)$  for some selected values of the
$\kappa$-parameter (where $z={A_{\kappa}}^{-1}$). Different from
Tsallis power-law functions which can become finite for some values
of the q-parameter,  the $\kappa$-distributions are not finite. In
other words, the thermal cutoff on the velocity space is not present
in such distributions regardless of the values of $\kappa$. It is
worth emphasizing, however, that this $\kappa$-distribution presents
the following mathematical behavior, viz.,
$f_{\kappa}(\eta)=f_{-\kappa}(\eta)$.

\section{Concluding Remarks}

In the last few years, several applications of the $\kappa$-power
law velocity distribution have been done in many disparate branches
of physics [8-21]. However, such investigations are usually related
with the $\kappa$-velocity distribution function as given by
equation (6).  On the other hand, many physical systems involve
naturally the presence of a conservative force field as happens for
example with ions in a magnetic field. Probably, the most popular
problem of a gas in a force-field is the planetary atmosphere. In
the standard simplified treatment, the temperature is uniform and
the tree-dimensional motion occurs under the action of a constant
gravitational field along the $z$-direction. To all this sort of
problems, the extended $\kappa$-distribution deduced here with basis
on the Vlasov equation, namely

\begin{eqnarray*}\label{eq18}
f_{\kappa}({\bf r},v) = A_\kappa\left[\sqrt{1
+\kappa^2\left(-\frac{m{\bf v}^2}{2k_BT} - \frac{U({\bf
r})}{k_BT}\right)^2} + \kappa \left(-\frac{m{\bf v}^2}{2k_BT} -
\frac{U({\bf r})}{k_BT}\right) \right]^{1/\kappa},
\end{eqnarray*}
can be applied, and, might prove to be of extreme wide
usefulness. Note also that a giroscopic term may also be added to
the above power law distribution. In a rotating frame  with constant
angular velocity, the whole effect is just to add a Coriolis term
$-1/2m\omega^{2}R^{2}$ to the potential energy, where $R$ is the
perpendicular distance from the axis of rotation. In the Newtonian
framework, such a term simulates a change in the potential energy
due to gravity. Finally, it is worth mentioning that the
present consistency between Vlasov equation and Kaniadakis
thermostatistics also is valid in the context of Tsallis
nonextensive framework \cite{1}.

\vspace{0.5cm}

\noindent {\bf Acknowledgments:} The authors are partially supported
by the Conselho Nacional de Desenvolvimento Cient\'{\i}fico e
Tecnol\'ogico (CNPq - Brazil). JASL is also grateful to FAPESP No.
04/13668-0.

\end{document}